\documentclass[9pt,sigconf,letterpaper,nonacm]{acmart}
\AtBeginDocument{
  \providecommand\BibTeX{{%
    \normalfont B\kern-0.5em{\scshape i\kern-0.25em b}\kern-0.8em\TeX}}}

\acmConference[ACM CoNEXT 2023]{The 19th International Conference on emerging Networking EXperiments and Technologies}{December 5--8, 2023}{Paris, France}
\usepackage{tikz}
\usepackage{amsmath}
\usepackage{graphicx}
\usepackage{filecontents}
\usepackage{lipsum}
\usepackage{tipa}
\usepackage{caption}
\usepackage{subcaption}
\usepackage{xspace}
\usepackage{amsmath}
\usepackage{hyperref}
\usepackage[shortcuts]{extdash}
\usepackage{soul}

\newcommand{\mm}[1]{}
\newcommand{\rv}[1]{}
\newcommand{\rvv}[1]{}
\newcommand{\id}[1]{}
\newcommand{\fc}[1]{}

\newcommand{\TOOL}{Sound-skwatter\xspace}

\newcommand{\graphemeTOphoneme}{\texttt{Grapheme\-/to\-/Phoneme}\xspace}
\newcommand{\ipaencoder}{\texttt{IPA\-/Encoder}\xspace}
\newcommand{\decoderTOgrapheme}{\texttt{Decoder\-/to\-/Grapheme}\xspace}
\newcommand{\decoderTOmel}{\texttt{Decoder\-/to\-/Mel}\xspace}
\newcommand{\postprocessor}{\texttt{Post\-/Processor}\xspace}
\newcommand{\duration}{\texttt{Duration\-/Predictor}\xspace}
\newcommand{\TOmel}{\texttt{to\-/Mel}\xspace}
\newcommand{\TTS}{\texttt{TTS}\xspace}

\usepackage{array}
\usepackage{makecell}
\usepackage[shortcuts]{extdash}
\begin{document}

\title[\TOOL (Did You Mean: Sound-squatter?)]{\TOOL (Did You Mean: Sound-squatter?) 
AI-powered Generator for Phishing Prevention}

% This is a tool for companies that want to search for spear phishing campaign using their name...

% We expect more impact of the feedback when we go to cross-language.


\author{Rodolfo Valentim}
\affiliation{%
  \institution{Politecnico di Torino}
  \streetaddress{Corso Duca degli Abruzzi, 24}
  \city{Torino}
%   \state{TO}
  \country{IT}
}
\email{rodolfo.vieira@polito.it}

\author{Idilio Drago}
\affiliation{
  \institution{Università degli Studi di Torino}
  \streetaddress{Via Giuseppe Verdi, 8, 10124 Torino TO}
  \city{Torino}
  \country{IT}}
\email{idilio.drago@unito.it}

\author{Federico Cerutti}
\affiliation{%
  \institution{Università degli Studi di Brescia}
  \city{Brescia}
  \country{IT}
}
\email{federico.cerutti@unibs.it}

\author{Marco Mellia}
\affiliation{
  \institution{Politecnico di Torino}
  \streetaddress{Corso Duca degli Abruzzi, 24}
  \city{Torino}
  \country{IT}
  \postcode{10129}}
\email{marco.mellia@polito.it}

\renewcommand{\shortauthors}{Valentim, et al.}

\begin{abstract}
% Your abstract text goes here. Just a few facts. Whet our appetites. 
% Not more than 200 words, if possible, and preferably closer to 150.

Sound-squatting is a phishing attack that tricks users into malicious resources by exploiting similarities in the pronunciation of words. Proactive defense against sound-squatting candidates is complex, and existing solutions rely on manually curated lists of homophones.

We here introduce \TOOL, a multi-language AI-based system that generates sound-squatting candidates for proactive defense. \TOOL relies on an innovative multi-modal combination of Transformers Networks and acoustic models to learn sound similarities. We show that \TOOL can automatically list known homophones and thousands of high-quality candidates.
In addition, it covers cross-language sound-squatting, i.e., when the reader and the listener speak different languages, supporting any combination of languages.

We apply \TOOL to network-centric phishing via squatted domain names. We find $\sim$ 10\% of the generated domains exist in the wild, the vast majority unknown to protection solutions. Next, we show attacks on the PyPI package manager,  where $\sim$  17\% of the popular packages have at least one existing candidate.

We believe \TOOL is a crucial asset to mitigate the sound-squatting phenomenon proactively on the Internet. To increase its impact, we publish an online demo and release our models and code as open source.

% 185 words currently 

\end{abstract}



% \ccsdesc[500]{Security and privacy~Phishing}
% \ccsdesc[500]{Computing methodologies~Neural networks}
% \ccsdesc[500]{Computing methodologies~Natural language generation}


\maketitle

\section{Introduction}

Cyber-squatting is a phishing attack that relies on the similarities of words to trick users into malicious systems, sites, or content. This type of identity attack has been applied in different contexts, from fake domains~\cite{holgers2006cutting, nikiforakis2014soundsquatting} and phishing campaigns~\cite{sonowal2019mmsphid, sonowal2020model}, to hijack the functionality of smart speakers~\cite{kumar2018skill, zhang2019dangerous} among others.
Several cyber-squatting strategies have been hypothesized and demonstrated in practice, including simple/frequent typos~\cite{wang2006strider, szurdi2014long, agten2015seven, khan2015every}, visual similarity of words~\cite{holgers2006cutting} and the collocation of common words~\cite{kintis2017hiding, loyola2020combo}.

%The way we interact with devices is dynamic and it is not restricted to screens anymore.

%The increase in the usage of voice controls to interact with devices (e.g., smartphones and smart speakers) changes the landscape for cyber-squatting attacks.
%Indeed, voice control is already popular thanks to the popularization of smart assistants, present on smartphones and smart speakers.

\emph{Sound-squatting} is a phishing technique that tries to trick users by leveraging words with pronunciation perceptively similar to targets. %This techniques tends to gain traction with the increase of voice controls usage.
Sound-squatting is generic and applies to the case of listeners that have to write (or interpret) a word pronounced by another person.
Whereas other types of cyber-squatting have been largely studied~\cite{tian2018needle, zeng2019squatting}, sound-squatting has received relatively little attention. %Interestingly, we found several well-known brands, e.g., Meta and Microsoft, that are aware of the vulnerability due to measures already taken against sound-squatting. 
Most previous studies are limited to the use of lists of known English homophones that lack coverage regarding the scope of such attack technique~\cite{nikiforakis2014soundsquatting}. 
Sound-squatting is challenging because it involves languages and multiple aspects that change across people.%For example, the way a word is understood varies considerably according to the mother tongue, accent, and voice characteristics of both the person pronouncing and the one listening to the word.

Current sound-squatting detection and mitigation strategies rely on manually curated lists of known homophones, i.e., two or more words having the same pronunciation but different meanings, origins, or spelling. For example ``new'' and ``knew''. In such lists, one can enumerate sound-squatting candidates, thus checking whether targets (e.g., domain names, internet profiles, etc.) may be a victim of tentative abuses~\cite{nikiforakis2014soundsquatting}.

This strategy has limitations. First, the set of homophones in a language is limited and it does not consider possibly not-existing words that might have perceptively similar pronunciations, i.e., quasi-homophones. Second, current solutions operate at a word level and do not consider character or syllable replacements and occlusions that have little effect on pronunciation. Third, homophones limit the sound-squatting generation to the same-language case. Being the Internet global, cross-language scenarios are likely popular. Here, the attacker exploits the inability of a person to correctly pronounce or write a word belonging to a foreign language.

%All in all, it opens to solutions for early mitigation of the phenomenon.

%to automatically generate words with similar pronunciations from some target has more coverage than static methods and is capable of producing more realistic names by performing the replacement at sub-word level, since names on the internet do not need to respect grammar rules.

We here introduce \TOOL, an AI-based system capable of \textit{automatically} creating sound-squatting candidates. \TOOL generates candidates for any given target name, working at the sub-word level and allowing configurable approximations during the search for candidates. \TOOL is built using a state-of-the-art Transformers Neural Network that can be trained for any language~\cite{vaswani2017attention}. For training, the network receives as input (i) the International Phoneme Alphabet (IPA) representation of the word and (ii) the spectrogram~\cite{shen2018natural} extracted from the target word pronunciation audio signal. At inference, it recreates the written form (\textit{grapheme}) while also considering  pronunciation. \TOOL uses a sequence-to-sequence model to find written alternatives with similar pronunciations.

%We advocate a data-driven approach to \textit{automatically} generate words with similar pronunciation, being them homophones, quasi-homophones, or not-existing words with similar pronunciation.\footnote{In the rest of the paper we refer to quasi-homophones also in the case the generated candidate is a non-existing word.} 
%A data-driven approach would come with several advantages.
%First, this approach would greatly increase coverage compared to methods based on homophones. Second, it produces candidates at the sub-word level, thus resulting better fitted to Internet domain names that do not necessarily respect grammar rules and syllables. Third, it can easily be ported to multiple languages, consider pronunciation and accent differences, and cover the cross-language cases that are today completely neglected. 

%In a nutshell, \TOOL includes both linguistic and acoustic features to the latent space mapping, thus informing the Transformers about sound similarities in different words.

% of the neural network computing several candidates for the same name target computing the candidate's probability of being a homophone.

We first validate \TOOL by comparing the list it generates against lists of known homophones. Here we find that \TOOL can automatically generated 84\% of known English homophones, and thousands of additional quasi-homophones, with the model with acoustic feedback that performs better than a simple system that does not include the audio part.

We show how \TOOL can help to mitigate domain name sound-squatting. We instrument it to generate automatically possible sound-squatting candidates of the top popular websites. Astonishingly, about 10\% of generated domains exists in the wild, but are unknown to current protection solutions. These include multiple cases we manually identify as malicious and on-sale/parking domains. Interestingly, some of these candidate domains are registered by the legitimate target owner as proactive protection against squatting.% All these results further corroborate the quality of the generated candidates.
%We show that our solution can be helpful to reveal several threats not detected by current solutions in domain sound-squatting.

We then check the existence of possible sound-squatting abuse in other domains. We focus on the PyPI repository for Python packages and we look for sound-squatting that target popular packages. \TOOL uncovers multiple cases. While these candidates do not seem malicious, the lack of structure and curation in the PyPI repository -- allowing such arbitrary names -- is a huge blind spot on the platform.

% We propose \TOOL, validate over a baseline of known homophones, discuss the cross-language homophones and apply to two different contexts: domain and PyPI repository sound-squatting.

% \footnote{The expression "previously unseen words" refers to words not necessarily are found in English vocabulary or that are not grammatically correct but that can be used as domain name, username or any name in the Internet.}.

In sum, our contributions can be summarized as follows:
\begin{itemize}
    \item We introduce \TOOL, an AI-powered sound-squatting automatic generator.% It relies on a Transformers Network enriched with a perceptual feedback from word pronunciations. \TOOL generates sound-squatting candidates at the sub-word level, increasing the coverage while allowing one to control the quality of candidates.
    
    \item We show that \TOOL lays the ground for cross-language sound-squatting search and mitigation. %Indeed, the Transformers architecture can be applied to reproduce the likely mistakes of a non-native speaker, e.g., generating candidates an American English speaker would expect when facing a foreign word.

    \item We provide a thorough analysis of \TOOL in two contexts, showing that sound-squatting is likely exploited for both domain squatting, and that PyPI packages.% to the attack.
\end{itemize}

\TOOL can be used for protection against attacks that target, e.g., brands. It results in an inexpensive solution for searching squatting campaigns in less controlled markets, i.e., automatically generating possible domain names that can be or are already abused by attackers.

To allow readers to test \TOOL in practice, we provide an anonymous running demo of the system online at \url{http://54.196.25.136/demo}. \TOOL pre-trained models, together with all our source code, will be available freely to the community upon publication.\footnote{Details are omitted to keep authors' anonymity. We suggest accessing the site from an anonymous IP address to avoid unveiling the reviewers.}

Next, we provide background information about sound-squatting and the neural network architectures we use in Section~\ref{sec:background}. We describe \TOOL in Section~\ref{sec:system_description} and validate it in Section~\ref{sec:validation}. We then discuss the results of domain name sound-squatting and PyPI packages sound-squatting in Section~\ref{sec:case_studies}. We discuss related work in Section~\ref{sec:related_work} before concluding the paper in Section~\ref{sec:conclusions}. 
\begin{table*}[!ht]
    \centering
    \caption{Examples of phonemes for American English with the grapheme and some examples (source \cite{dyslexia2022}).}
    \label{tab:examples_of_phonemes}
    \begin{tabular}{c|c|l|l}
        Type      & Phoneme     & Grapheme                                         & Examples \\ \hline
        Consonant & \textesh	& sh, ce, s, ci, si, ch, sci, ti                    & sham, ocean, sure, special, pension \\ \hline
        Consonant & v           & v, f, ph, ve                                      & vine, of, stephen, five \\ \hline
        Vowel     & æ	        & a, ai, au                                         & cat, plaid, laugh \\ \hline
        Vowel     &	e\i	        & a, ai, eigh, aigh, ay, er, et, ei, au, ae, ea, ey & bay, maid, weigh, straight, pay \\\hline
    \end{tabular}

\end{table*}

\section{Background}
\label{sec:background}

\subsection{Cyber-squatting and sound-squatting}
\label{sec:cyber-squatting}

Cyber-squatting is a class of attack in which a malicious actor tries to impersonate a legitimate resource \cite{CAPEC-616}. Cyber-squatting has gained notoriety with the widespread deployment of domain-squatting, a type of attack in which attackers register fake domain names to divert traffic from popular websites. For example, an in-depth search of over 224 million DNS records in 2018 identified 657\,k domain names likely impersonating 702 popular brands~\cite{tian2018needle}.

The Mitre Att\&ck's CAPEC-631~\cite{CAPEC-631} defines sound-squatting uniquely in the context of domain-squatting, as an attack in which \textit{``an adversary registers a domain name that sounds the same as a trusted domain, but has a different spelling''}.
However, sound-squatting is more generic than that. For instance, abuses in Alexa voice assistant have already been identified~\cite{kumar2018skill}. In general, the sound-squatting attack tries to diverge users into malicious resources based on the assumption that users (or devices) will confound words with similar pronunciations \--- e.g., clicking on links, typing in similar malicious words, or misunderstanding an action.

Considering only domain-squatting, the CAPEC-631 also enumerates possible mitigation to sound-squatting: (i) the deployment of additional checks when resolving names in the DNS, together with the authentication of servers, and (ii) the preventive purchase of domains that have potential for sound-squatting.
%We will show in Section~\ref{sec:case_studies} that major brands use the latter as a protection mechanism.
In addition to these protections, the system can warn the user about possibly malicious use of a word. 

In all cases, the targets of sound-squatting must know the names attackers will use to impersonate their brands. \TOOL comes precisely to solve this problem, generating candidate names \--- ranked with their probabilities \--- that can be used to mitigate the problem proactively.

\subsection{International Phoneme Alphabet (IPA)}
\label{sec:phonemes}

Continuous speech is perceived in segments. These segments are classified according to how the vocal tract produces them. Each natural language uses a different set of segments. Phonemes are the smallest units in a language to distinguish word meanings. Phonemes are precise phonetic realizations often affected by social variation, regional dialect, etc. The International Phonetic Alphabet is a means to represent such phonemes with standard symbols (IPA)~\cite{internationalphoneticassociation2022}.

The International Phonetic Association introduced IPA in the $19^{th}$ century. IPA uses Latin and Greek script symbols, transcribing sounds (phones). Letters are organized into categories, i.e., vowels and pulmonic/non-pulmonic consonants. Some examples are shown in Table~\ref{tab:examples_of_phonemes}~\cite{dyslexia2022}.

Table~\ref{tab:examples_of_phonemes} shows that some phonemes may be equivalent to more than one grapheme. This relation varies from language to language, with some languages more \emph{phonetically consistent} than others. Italian and German are two examples of languages with high consistency, i.e.,
most graphemes have a one-to-one relation with a phoneme. English and many other languages are not phonetically consistent, largely missing such a direct relation. Languages with considerable inconsistency are more prone to confusion when interpreting/writing a pronounced word. Inconsistent languages are thus particularly vulnerable to sound-squatting.

\TOOL uses IPA to encode the input words. There exist solutions that translate any word in the corresponding IPA. For instance 
the  eSpeak NG (Next Generation) \TTS engine\footnote{\url{https://github.com/espeak-ng/espeak-ng}} supports more than 100 languages. \TOOL uses it to derive the input IPA given a target word and language.

\subsection{Transformers Neural Networks}
\label{sec-transformers}

Transformers are a hot topic in the deep neural network community. Initially proposed for translation~\cite{vaswani2017attention}, they have been used as the means to achieve natural language processing (NLP), computer vision (CV), and speech processing.

The vanilla implementation of Transformers consists of an encoder and a decoder, each of which is a stack of N identical blocks. Each encoder includes a MultiHead Attention block and a feed-forward interconnected by \textit{Add and Normalization} layers. The decoder is similar to the encoder; however, it adds to the encoder's architecture a cross-multihead attention mechanism that computes the attention between the input and the previous states of the target.

An important aspect to discuss regarding the Transformers is the difference between the training and the inference. During training, the Transformers use a mask over the target states given as input for the cross-attention head. This mask allows the model to learn multiple states simultaneously. 

The inference process happens in steps where the decoder receives the previous states generated to compute the cross-attention. In this way, the output of the next element depends on the sequence of the previously generated elements. This is fundamental to correctly predict the next character in a word or the next word in a sentence. %For this reason, it is said that the Transformer is non-autoregressive at the training and autoregressive  at inference.

\TOOL uses Transformers to generate words at the character level. It considers not only the most probable character but also explores possible words that may derive from considering the top-K most probable following characters, generating possible alternative ways to write the same input IPA sequence.

\subsection{Mel Spectrogram}
\label{sec-MEL}

The spectrum describes a signal's magnitude and phase characteristics as a function of frequency~\cite{SEMMLOW2018245}. A spectrogram is a visual representation of an audio signal decomposed in frequencies where one axis represents time, and the other represents frequencies. 

\begin{figure}
    \centering
    \includegraphics[width=0.8\linewidth]{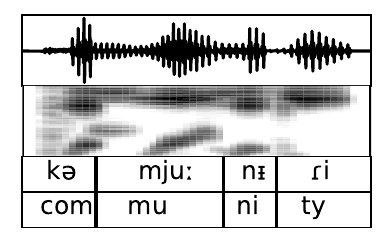}
    \caption{An raw audio wave and its Mel Spectrogram aligned to the pronunciation and grapheme of the word \textit{``community.''}}
    \label{fig:audio_example}
\end{figure}

The Mel Spectrogram is a log-scaled version of a Linear Spectrogram.
%The Mel Scale uses a logarithmic transformation of a signal's frequency. 
This representation is more suitable for humans because it is easier for our auditory system to distinguish between similar low-frequency sounds than between similar high-frequency sounds. 
The audio representation in Mel Spectrogram is, therefore, often used in machine learning applications because it helps to focus on the components of frequencies that matter for speech applications. 
Figure~\ref{fig:audio_example} shows the raw audio signal, the Mel Spectrogram, the pronunciation in IPA, and the grapheme for the word \textit{``community.''} 

\TOOL uses the Mel Spectrogram to learn the representation of a sound and uses the representation to generate graphemes whose pronunciation is close to the original sound. \TOOL also uses the alignment of the pronunciation with the signal to infer the duration of each phoneme, as shown in Figure~\ref{fig:audio_example}. Such duration is pivotal also for reconstructing the spectrogram, as discussed in more detail in the following section. %Both data allow the Transformer Network to embed also acoustic models that learn sound similarities.
\section{\TOOL: System Description}
\label{sec:system_description}

\begin{figure*}[ht]
 \centering
  \begin{subfigure}[b]{0.475\textwidth}
     \includegraphics[width=\textwidth]{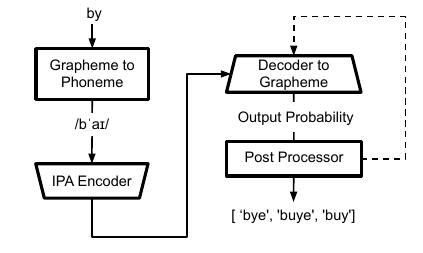}
     \caption{Architecture used during inference.}
     \label{fig:inference}
 \end{subfigure}
  \hspace{0.2cm}
  \begin{subfigure}[b]{0.475\textwidth}
     \centering
     \includegraphics[width=\textwidth]{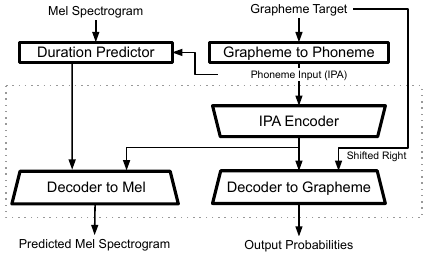}
     \caption{Architecture used during training.}
     \label{fig:training}
 \end{subfigure}

\caption{%
(\subref{fig:inference}) The process to generate candidates is composed of an \ipaencoder that maps the input to a latent space and a \decoderTOgrapheme that reconstructs the input.
%and a \decoderTOmel that renders the pronunciation in frequency space at the Mel scale.
% 
(\subref{fig:training}) The training architecture for learning how to generate quasi-homophones. The dotted box highlights the components that are trained for the generation of quasi-homophones: \duration is a pre-trained function. It receives as input the phoneme translation of a word and the duration of each phoneme in the expected spectrogram and outputs a reconstructed spectrogram and the probabilities that are used to find the grapheme translation.
}
\end{figure*}

\TOOL is an AI-based system capable of creating sound-squatting candidates that relies on Transformers Neural Network to translate from phonemes to graphemes.
We start by describing the part of the architecture used during inference.
It operates as depicted in Figure \ref{fig:inference}:
\begin{enumerate}
    \item \graphemeTOphoneme \--- built upon the  eSpeak NG (Next Generation) \TTS engine\footnote{\url{https://github.com/espeak-ng/espeak-ng} (visited on 10/10/2022).} \--- transforms an input word written in grapheme form (e.g., the word \texttt{by}) into its IPA representation (e.g., \texttt{/b\textprimstress a\textsci /}). %\TOOL could be trained to use any languages in grapheme representation;
    \item \ipaencoder then encodes such an IPA word into a vector representation.

    \item \decoderTOgrapheme decodes the vector representation into words written in grapheme form which are quasi-homophones to the input word (e.g., \texttt{bye}).

    \item  \postprocessor selects the quasi-homophones to be presented to the user.
\end{enumerate}

% \ipaencoder and \decoderTOgrapheme are discussed in details in Section \ref{sec:encoder-decoder}.
% The actual selections of quasi-homophones is left to a \postprocessor further discussed in Section \ref{sec:postprocessor}.

The architecture used to train the \ipaencoder and \decoderTOgrapheme \--- the core of our proposal \--- is depicted in Figure \ref{fig:training}. Both modules are trained jointly with one further function, \decoderTOmel, exploiting multi-modal data coordinated by the \duration function. The overall training task thus:
\begin{itemize}
    \item is analogous to a sequence-to-sequence model, where the \ipaencoder and \decoderTOgrapheme together translate from IPA to the target language grapheme. At the same time the Mel Spectrogram supervisory signal is reconstructed from the phoneme vector representation and the forecast duration of each phoneme in the spectrogram domain;
    \item uses as input pairs $\langle x, \TOmel(\TTS(x)) \rangle$, where $x$ is a word in phoneme format, and \TOmel and \TTS are, respectively, one of the many existing software that generates a Mel Spectogram of audio signals and Text-to-Speech software, eventually first calling the \graphemeTOphoneme function for having it in IPA. The dual-modality input is required for generating quasi-homophones of good quality;
    \item outputs the word in grapheme format (via \decoderTOgrapheme) and the Mel Spectogram (via \decoderTOmel) from the vector representation which \ipaencoder outputs. The loss function used for training needs to carefully balance the training process between these two modalities.
    The loss is computed by summing the $L1$ and $L2$ Loss of the Mel Spectrogram reconstruction and the Cross-entropy Loss calculated between the predicted and the expected grapheme.
\end{itemize}

\duration, a pre-trained function, coordinates the flow of multi-modal data by feeding the predicted temporal duration of each of the phonemes in $\graphemeTOphoneme(x)$ to \decoderTOmel.
It uses Convolution and LSTM Units to minimize the Connectionist Temporal Classification (CTC) loss \cite{Graves2006}, i.e., a loss between a continuous time-series \--- the Mel Spectrogram \--- and a target sequence \--- the IPA word. The former is obtained  by applying the Fast Fourier transform (FFT) over windowed segments of the audio signal and a transformation to Mel Scale. Given $K$ the length of $x$, our word in phoneme format, \duration outputs $\langle d_1, d_2, ..., d_K\rangle$, where $d_i$ represents the number of windowed segments of the audio signal predicted to be associated to the $i$-th phoneme of $x$.

\TOOL leverages Transformers Neural Network to train \ipaencoder, \decoderTOgrapheme, and \decoderTOmel (within the dotted box of Figure \ref{fig:training}). Transformers rely on the
%The Transformer models uses
self-attention mechanism to learn how to translate a word in IPA back into grapheme format. The output of \ipaencoder is forwarded to \decoderTOgrapheme, which uses the contextual information extracted from the input by \ipaencoder, together with the current state of the output. As discussed in Section~\ref{sec-transformers}, a Transformer is an auto-regressive model at inference, i.e., it can leverage its own history of predictions to forecast future states.

% The \duration module is a model, trained independently and prior to the \TOOL. We use Convolution and LSTM Units to find and alignment between the IPA sequence and the Mel Spectrogram that minimized the Connectionist Temporal Classification (CTC) loss. It produces an output $D = [d_1, d_2, ..., d_n]$ for $n = [0, length(x)]$ which is corresponds to the number of Mel Spectrogram window steps for each phoneme. \rv{@federico, sorry for the freestyle math notation. Can you fix if it's wrong?}

In particular, \decoderTOgrapheme, given an IPA input and the predictions made for each of the previous $N-1$ characters, forecasts the probabilities  of each possible character being the $N$-th character.
% \decoderTOgrapheme predicts one character at a time, and it requires knowing the predictions of all the previous $N-1$ characters for forecasting the $N$-th character.
% In practice, the output of \decoderTOgrapheme is the probability of each possible character being the next character.
It is then the role of \postprocessor to look at these probabilities and feed the history back to the \decoderTOgrapheme for the new forecast, as discussed
in more detail in the following.

\subsection{\postprocessor: the Quasi-Homophone Generation}
\label{sec:postprocessor}

The \postprocessor is the last element of the inference: it receives as input the \decoderTOgrapheme's probabilistic forecasts of the next character, and it keeps track of the history of predictions to feed back to \decoderTOgrapheme.
Because \decoderTOgrapheme operates as an auto-regressive model at inference, by tweaking the history the \postprocessor feeds back, we can generate more than a candidate quasi-homophone. We use a Beam Search algorithm to find the best candidates. 

\begin{figure}[t]
    \centering
    \includegraphics[width=1.1\linewidth]{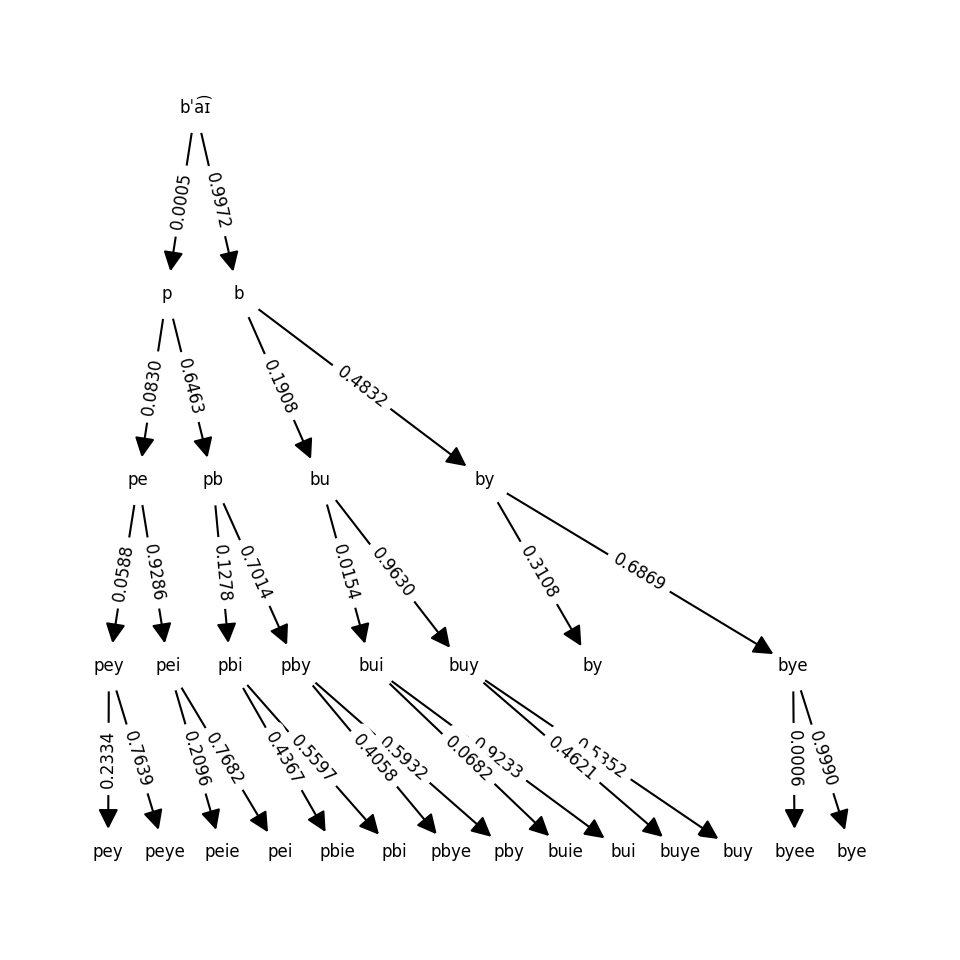}
    \caption{Illustration of the inference process with $K=2$ children per node. At each inference step, we collect the two next characters with the highest probability. The left and the right children have the second-highest and highest probability, respectively. Some nodes do not have children because they reach the "End of Sentence" state.}
    \label{fig:binary-inference}
\end{figure}

%
%We hypothesised that it is possible to produce more than one grapheme for some pronunciation. However, in other to perform this task, we need to tweak the inference process. The process can be described as a binary tree.
%
At each inference step, the \postprocessor stores the $C$ most-likely predictions of the \decoderTOgrapheme, and it constructs alternative histories to feed back to \decoderTOgrapheme at the following inference step. Figure~\ref{fig:binary-inference} depicts this iterative process in the form of a tree (with $C=2)$, where each edge has an associate probability. It starts from the IPA representation of the word \texttt{by}, and each node represents the two most likely next-character given the past sequence of characters. The letter \texttt{b} is the best candidate first letter (with probability 99.72\%), while \texttt{p} results the second-highest. \texttt{b} can then be followed by \texttt{u} or \texttt{y}, etc. After repeating the exploration four times, the process generates 15 different ways to write \texttt{by}. Notice that the \postprocessor stops the generation when the \decoderTOgrapheme outputs the special character \texttt{EoS} (End of Sentence). This happens for instance when generating \texttt{by}.

% which are At each inference step we select the two highest probabilities for the next state. With means, we collect the two most probable characters that can produce the same pronunciation. At this step duplicate the current status. This is the branching in the tree. The two different states are fed to the decoder with the same pronunciation. The current state together with the pronunciation guides the next state.
% We show an example for the word "by" in Figure~\ref{fig:binary-inference}, we show the exact output for 4 iterations.

%As seen, there is a probability associated to every edge.
To find the best quasi-homophone candidates, \postprocessor computes the joint probability of each leaf as being the product of the edge probabilities. At each step, \postprocessor ranks current leaves by joint probabilities and keeps the top-$K$ most likely ones to prune the search space and avoid pursuing branches with a low likelihood of producing good quasi-homophones. The number of iterations $M$ at which stop the generation process, the number of candidate predictions (children) $C$ to generate at each step, as well as the number of best candidates $K$ to keep are additional hyper-parameters of \TOOL which can be either defined manually or identified using standard local-search procedures.
%
% The number of how many candidates to keep is a choice. In this task, an static or proportional threshold is hard to define, since the probabilities greatly vary from word to word.
%
For the purpose of this work, given $N$ the length of the input word, \postprocessor iterates $M=N+6$ times, generating potentially $C^{N+6}$ candidates. To limit the exploration, we set $K=64$ and choose $C=2$. These parameters have been determined by manual inspection, and it is clearly heavily dependent on the tasks and datasets of interest.

\section{\TOOL: Training and validation}
\label{sec:validation}

After defining the model for generating homophones and quasi-homophones, we describe the language to which we will apply our model. We select American English since it is the most used language for digital services and on the Internet. English is also phonetically inconsistent, thus a potential target for squatting. Recall, however, that the model and methodology are language-independent, and \TOOL can be trained for any language. We collect a training set of words, IPA pronunciation, and Mel Spectrogram to train \TOOL (Section \ref{sec:dataset}). We then validate the trained \TOOL using a list of known homophones  and check if it can generate them (Section \ref{sec:validation_results}). %In the following, we detail both processes of training and validation.

\subsection{Dataset and training}
\label{sec:dataset}

The dataset used during training contains American English words. We consider the list of words from the GNU Aspell~\cite{atkinson2006gnu} reference word list. GNU Aspell is a free and open-source spell checker and contains $125\,929$ words in total for American English.

To acquire the pronunciation, we use the open-source software eSpeak NG Text-to-Speech. It supports more than 100 languages, and we set it for American English to solve the grapheme-to-phoneme translation and to produce the speech sound signal. The generated speech sounds artificial to a human ear, but it is adequate for our goals.%is not smooth and natural and is clearly software generated. However, for our proposal it is sufficient.

The final step is to convert the sound signal to a spectrogram at the Mel Scale. We use Torch Audio library~\cite{yang2021torchaudio} to produce those.

We train the model with a batch size of 64 words. The training set contains $80\%$ of the samples, and we preserve $10\%$ for both the validation and the test sets. We use the validation set to choose the best model and the test set to verify overfitting. We use the Adam optimizer with $LR = 0.0001$, $\beta_1 = 0.9$, $\beta_2 = 0.98$, $\epsilon = 10^{-9}$. We set the step learning rate decay with $\gamma = 0.1$ scheduled every $10$ epochs. We train the model for $30$ epochs (about 47k steps) which took around $168$ minutes on a single Nvidia Tesla v100. We train \duration with the same dataset and the same Adam optimizer. \duration converges at around 100k steps. Further details regarding the training are found in Appendix~\ref{ap:details_training}.

\subsection{Validation}
\label{sec:validation_results}

We verify if the model can generate homophones by selecting known sets of homophones for American English. We consider as homophones any group of words having the same IPA encoding. Then, we choose one word for each set and ask \TOOL to generate candidate squatting words. We expect a considerable intersection between the set of known homophones and the set of generate candidates.

\begin{table*}[t]
    \caption{Evaluation of the capability to generate homophones using as control the known homophones in American English. The last column contains the candidates ordered by the probability of being correct from the highest to the smallest. Notice that the known homophones are at the beginning of the list because they are, in fact, the best candidates for the given pronunciation.}
    \label{tab:model_evaluation_results}
    \centering
    \begin{tabular}{c|l|l}
        \textbf{Word} &          \textbf{Known homophones} &                                 \textbf{Candidates Ordered by Joint Probability} \\ \hline
        hobbes &             hobbs, hobs & \textbf{hobbs}, \textbf{hobs}, haabs, hobbes, haabs, hobbes, ho... \\ \hline
        niche &         nitsche, nitsch &  nitch, \textbf{nitsch}, nich, \textbf{nitsche}, knich, niche, k... \\  \hline
        gainer &  gaynor, gayner, gainor &  gainer, \textbf{gainor}, \textbf{gayner}, \textbf{gaynor}, gainer, gayne... \\
    \end{tabular}

\end{table*}

For a set of $125\,929$ words in American English, we found $2\,279$ sets of homophones for a total amount of $5\,804$ homophones. In the validation process, we generate $47\,024$ candidates: on average, we produce $18.49$ candidates per target, with a standard deviation of $7.61$. The intersection size between the candidates and the known homophones is $4\,870$, i.e., \TOOL found $83.91\%$ of the possible homophones. We consider this a very good coverage, allowing us to conclude that \TOOL can generate words with exact and similar pronunciation from a target. We show some examples in Table~\ref{tab:model_evaluation_results}.
In addition, \TOOL generates possible alternate spelling of the same target work, introducing several degrees of difference. Candidates are ordered according to the joint probability. Notice that homophones (in bold) appear among the top candidates. 

Checking if the generated words are existing in the American English vocabulary, we find that out of $47\,024$ candidates, $1\,975$ are actual words, i.e., they are quasi-homophones.\footnote{We use \href{https://linux.die.net/man/1/enchant}{Enchant Spellchecker} dictionary to verify if the word exists in the US English vocabulary.} The remaining words are not present in the dictionary but show certain similarities with the target word.

\begin{figure}[t]
    \centering
0    \includegraphics[width=\linewidth]{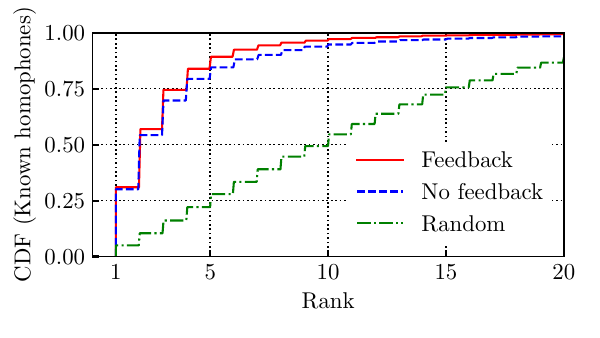}
    \caption{The ECDF for the rank in which we find the known homophones on the list of generated candidates when sorted by joint probability. As the ECDF is substantially higher than random ordering, we can conclude that using the joint probabilities is a valid method.
    %We compare the generation with a random ordering.
    }
    \label{fig:comparisson_generation_with_random}
\end{figure}

To confirm that the joint probability is a valid method to rank the candidates, Figure~\ref{fig:comparisson_generation_with_random} shows the Empirical Cumulative Distribution Function (ECDF) curve that represents the probability of finding a homophone in the top $n$-th elements in the list. The red curve refers to sorting by the joint probability. The green curve refers to random sorting. Notice that more than $80\%$ of the homophones are found in the top 5 candidates, a substantially higher proportion than a random choice.

\subsection{Impact of acoustic feedback}
Here we compare \TOOL with a simpler model where we omit the \duration and the \decoderTOmel. While simpler, it is also limited in the cross-language case since the audio part plays an important role in such case.

Comparing the results on the known English homophone test, we observe that \TOOL generates $4\,870$ with sound feedback and $4748$ without. The different is $122$. However, we see a bigger difference when we compare the unseen homophones and quasi-homophones, with additional $938$ words found including the acoustic feedback.

For completeness, we report in Figure~\ref{fig:comparisson_generation_with_random} the ECDF curve that represents the probability of finding a homophone in the top $n$-th elements in the list for the model without acoustic feedback. The curve for the model without feedback in blue is to the right of the red curve indicating a lower quality in the generation because the known homophones are found later in the rank of generated candidates.

% \begin{table}[]
%     \centering
%     \begin{tabular}{c|c|c}
%         \textbf{Feature} & \textbf{With} & \textbf{Without} \\ \hline
%         Number of words & 2279 & - \\  \hline
%         Known homophones & 5804 & - \\ \hline
%         Intersection  & 4870 & 4748 \\  \hline
%         Unseen candidates & 42154 & 41216 \\
%     \end{tabular}
%     \caption{Caption}
%     \label{tab:my_label}
% \end{table}

Some examples are shown in Table~\ref{tab:with_without}. We notice that the generated samples using the feedback capture more subtle sound similarities, which is good for sound-squatting applications. For instance, without sound feedback, the model it is not capable to assign to the word 'silvester' homophones ending in with the syllable 'tre' as 'silves\textbf{tre}', and 'sylves\textbf{tre}'. This is a subtle similarity that the acoustic feedback enables during the generation. The addition of acoustic features bring a qualitative improvement, without performance penalties.

\begin{table*}[]
\centering
\caption{Evaluation of the increment of candidates added by the increment of a sound feedback to the model. Preliminary results show that we have an increase of variability in the generation.}
\label{tab:with_without}
\begin{tabular}{c|l|l|l}
\textbf{Word}      &                           \textbf{Known homophones} &    \textbf{Gen. With Sound Feedback} & \textbf{Gen. No Sound Feedback} \\ \hline
pauley    &                      pauli, pawley, paulie &            paulie, pauli, pawley &                           pawley \\ \hline
chae      &  quaye, kea, ke, quay, key, keye, khe, kee &      kee, key, quaye, keye, quay &             ke, kea, quay, quaye \\ \hline
ais       &                 uys, eis, eyes, ise, ayes  &                   ise, eyes, eis &                              ise \\ \hline
heines    &         heinz, heinze, heins, hines, hynes &             heinze, heinz, hines &                     heins, hines \\ \hline
silvester &            sylvester, silvestre, sylvestre &  silvestre, sylvester, sylvestre &                        sylvester \\

\end{tabular}
\end{table*}

\section{Case studies}
\label{sec:case_studies}

% \mm{move to intro or so \TOOL is capable of finding homophones and words with similar pronunciation for previously unseen words. The concept of using homophones to lure uses into unwillingly release sensitive information is known as sound-squatting. This technique is majorly used for phishing domain scams. }

In this section, we explore possible scenarios where sound-squatting may be a threat and show how \TOOL can be used to prevent abuses.
Our goal is to demonstrate that \TOOL can be used for proactive protection in any context that requires names as identifiers.
We do not intend to present an extensive study on the possible abuse of sound-squatting in the wild, and we limit to illustrate that this type of attack is possible and \TOOL can help to mitigate it.

Armed with a list of candidates, one can apply standard solutions to protect the original target. These solutions include blocklisting, preventive registrations, showing warnings to the users, etc.

\subsection{Same- and Cross-language sound-squatting}

\begin{table*}[]
    \caption{Examples of same- and cross-language sound-squatting for different languages. The possible pronunciation approximations that \TOOL generates are ordered according to their joint probability.}
    \label{tab:examples_cross_squatting}
    \centering
    \begin{tabular}{c|c|c|l}
        \textbf{Language $A$} & \textbf{Language $B$} & \textbf{Grapheme} & \textbf{Cross-language homophones} \\ \hline
        US English & US English & community & community, komunity, kommunity, comunity, kemunity, chommunity, ... \\ \hline
        US English & Italian & comunità &  communita, komunita, communeta, comunita, kemunita, kommunita, ... \\ \hline
        US English & Portuguese & comunidade & communidade, communidayed, communidayde, communedayed, ... \\ \hline
        US English & Spanish & comunidad & communidad, communedad, communidade, komunidad, komuniadad, ...  \\
    \end{tabular}

\end{table*}

When asking \TOOL to generate a homophone, we have considered the case of same-language sound-squatting, i.e., the legitimate and the sound-squatting candidates derive from the same language. This is the case of, e.g., an English speaker that has to write a word while listening to an English reader uttering the word.

Cross-language sound-squatting explores the pronunciation similarity between words in different languages instead. Confusion might occur in any combination of languages and can be a means of creating effective sound-squatting phishing campaigns. The attack can be highly effective if attackers focus on regions (and groups of people) where brand owners neglect protection measures. 

% The difficulties of generating and detecting such an attack rely in the need of understanding any pair of languages to be exhaustive.

Cross-language squatting might happen in different situations, such as:

\begin{itemize}
    \item \textbf{Someone reads a foreign word to a second person who writes it in their own language}. A speaker of a language $A$ reads some word from a language $B$, pronouncing the grapheme according to language $A$ rules and understanding/transcribing the word wrongly.
    \item \textbf{Someone listens to a foreign word and writes it in her native language spelling}. A speaker of a language $A$ listens to the pronunciation of some word from a language B and phonetically tries to write the word in her own language;
\end{itemize}

%An attacker could leverage differences in the languages to lure users.

Consider, for example, the Portuguese word \textit{``milho.''} In the first case, \textit{``milho''}, when read by an American English speaker, would be similar to \textit{``millhoe.''} %Using a voice assistant, \texttt{milho.com} could then become \texttt{millhoe.com}.
This attack can be used together with smart-speakers and Text-to-Speech tools too, where the word is miss-pronounced and induce the device to make a mistake, e.g., trick a smart-speaker in: ``Alexa, access to \texttt{millhoe.com}'' but, actually the user wants to go to \texttt{milho.com}.
To mimic this type of attack, \TOOL needs to get as input the foreign language word grapheme form. The \graphemeTOphoneme would ``read'' it as if it was English, and the rest of the architecture would generate possible grapheme forms an English writer could produce.

In the second case, the \textit{``lh''} in Portuguese sounds like the English \textit{``ly.''} An English listener listening to the Portuguese word would understand the pronunciation as something like ``mee-lee-yo.'' An attacker could thus create this cross-language homophone \textit{``meeleeyo''} and use it to trick users into accessing to \texttt{meeleeyo.com} instead of \texttt{milho.com}.
%if language B is Portuguese and language A is American English.

In \TOOL, the \graphemeTOphoneme is configured to read the input word in Portuguese, giving its IPA representation as output. The \ipaencoder, trained with English, will then interpret the IPA phonemes as an English listener, and the \decoderTOgrapheme would generate possible English grapheme forms for the target word.
Notice here that the \ipaencoder must be able to handle the case of phonemes that are not present in the destination language but in the original language. A simple replacement with the closest phoneme policy would suffice. 
%We leave this case for future work on how to generalize this.

We show some examples of how \TOOL can automatize the creation of cross-language squatting attacks of the first case. We train \TOOL to emulate how an American English reader would read and then write the grapheme form of words she reads from a foreign language. 

%Although we do not explore the second case in this work, we believe \TOOL can be adapted by changing the language used in \graphemeTOphoneme to perform the translation from grapheme in Language B to phoneme in Language B.
% In this configuration, \TOOL can replicate how an English American speaker would read an world in a foreign language and how would they write such pronunciation.

As language B, we select three Latin languages: Italian, Portuguese and Spanish. Consider the word \textit{``community.''} It has similar grapheme forms for all languages \--- see Table~\ref{tab:examples_cross_squatting}. In this experiment, we ask \TOOL to generate homophone candidates in language $A$ mimicking how a language $A$ speaker would pronounce the original word in language $B$.

Table~\ref{tab:examples_cross_squatting} lists examples of words that might induce a user to errors.
The possible pronunciation approximations that \TOOL generates are ordered according to their joint probability. The first line shows the same-language squatting for completeness, i.e., in American English. 

Here \TOOL outputs the correct form to write ``community'' as the first word, which is the one with the highest probability. Then it proposes a list of possible candidate homophones.
For Italian, the correct way of writing ``comunità'' is not even present in the list, i.e., \TOOL mimics an American English reader that would likely not be able to write such a word correctly. The same happens in other languages.

In the following, we use \TOOL to automatically generate candidates starting from words taken from real use cases. We consider the domain name squatting and Python packages sound-squatting, i.e., someone registering domains that may sound similar to a given target domain or submitting Python packages with similar names to popular packages.

\subsection{Domain Name Sound-Squatting}

Domain squatting is a technique that attackers use when they register perceptively confusing domain names aiming at tricking visitors into them~\cite{zeng2019squatting}.
% There are several types of domain squatting: typo-squatting, bit-squatting, homograph-squatting, sound-squatting, combo-squatting~\cite{zeng2019comprehensive} and skill-squatting~\cite{kumar2018skill}. Each type explores one perception aspect to lure users into the false domains.
%
Among the different squatting types, domain squatting based on sound-squatting has received little attention while gaining traction with the advent of smart speakers and voice-assistants~\cite{kumar2018skill}. %Sound-squatting explores the pronunciation of domains and the fact that different words might present the same sound even if written differently. Sound-squatting is challenging because pronunciation varies from language to language and user to user, which may introduce different errors when typing or speaking a word. The state-of-art relies on statically built lists of homophones, which however is hard to generalize~\cite{zeng2019squatting}.
%\id{all this is said in the intro} \rm{yes - we can remove it}

Here we present a thorough analysis for domain sound-squatting using \TOOL to produce quasi-homophones. Given target domains, we generate sets of squatting candidates and check if such domains exist. The process is as follows: (i) we collect a list of domains from target services, (ii) we extract the Second Level Domain names, as the portion of the domains susceptible to abuse; (iii) we run the domains over \TOOL and collect the generated candidates. \TOOL is trained as an American English speaker listening to the domains in English and writing them in English. If the original target domains are English words, \TOOL generates same-language candidates. If the original domain is of a different language, \TOOL generates cross-language candidates.

Given the list of candidates, we verify the quality of our generation by assessing if any of these domains exist. The process is as follows: (i) We form fully qualified domain names by concatenating the quasi-homophones to a Top Level Domains (TLDs); (ii) We verify if such domains are actually registered; (iii) If a domain exists, we query external sources to classify the domain as \textbf{Malicious} or \textbf{Suspicious}. For this last step, we use Google Safe Browsing to mark the domain as Malicious and Virus Total to flag it as Malicious or Suspicious.\footnote{Virus Total uses a reputation mechanism to flag domains as Suspicious or Malicious.} If the domain results are not present in any lists, we check if it could be a \textbf{Target-Owned} domain, i.e., a domain registered by the same company that owns the target domain. This protection mechanism is usual for companies and brands concerned about their reputation. Usually, these Target-Owned domains redirect the user to the correct domain when accessed. To detect if the candidate domains belong to the Target-Owned class, we compare the information registered in the \texttt{Whois} database for the target and candidate domains. If the owner is the same, and it is a first-party identifiable from the brand (i.e., not a third-party service), we declare the candidate domain as Target-Owned. Finally, we classify as \textbf{Unknown} the remaining candidates.

% \id{put more precise comment about the whois case, it should be clear first-party, non third-party ``anonymous'' owner}

\subsubsection{Experiment design and results}

For this experiment, we select the 200 most popular websites from the Similar Web list~\cite{similarweb}. This list includes domains in any language, thus leading to both same- and cross-language cases.
After extracting only the Second Level Domains, we have 164 unique words that we consider as our targets.\footnote{We handle the case of meaningless second-level domains, e.g., \texttt{.co.uk}} We generate candidates for each of them, asking \TOOL to output at most the $K=64$ most likely domains according to their joint probability. As shown in Table~\ref{tab:summary_first_domain_analysis}, in total we obtain $3\,681$ candidates,\footnote{Naturally, this number excludes the original target name, if \TOOL outputs the target as illustrated in Table~\ref{tab:model_evaluation_results}.} an average of $22.44$ by target, with standard deviation of $11.60$.

We next consider the 10 most popular TLDs again according to Similar Web. They are: $.com, .ca, .org, .ru, .net, .au, .uk, .in, .ir,$ and $.de$.
The product between any target and any TLD produces $36\,810$ candidate domains, out of which $3\,692$ results are registered. This number represents $10.03\%$ of the total.

\begin{table}[]
    \caption{Summary of sound-squatting for Domain names.}
    \label{tab:summary_first_domain_analysis}
    \centering
    \begin{tabular}{c|c}
        &  \textbf{Quantity} \\ \hline
        Initial Domains & 200 \\ \hline
        Unique Domains & 164 \\ \hline
        Generated Candidates & 3\,681 \\
        %& $Mean=22.44$ \\ & $SD=11.60$ \\
        \hline
        Candidate Domains & 36\,810 \\ \hline
        Existing Candidates & 3\,692 (10.03\%) \\
    \end{tabular}
\end{table}

We classify the existing candidates by first searching at Virus Total and Google Safe Browsing blocklist. The Table~\ref{tab:summary_second_domain_analysis} summarizes the results. The search on Google Safe Browsing offers surprisingly lacking results with only 4 candidates being present in the blocking list, i.e., telegram: \texttt{telegragm.com}, shopee: \texttt{shhopee.com}, as: \texttt{az.org}, cnbc: \texttt{snbc.com}. All the other 3\,677 candidate domains are not present.
The search on Virus Total shows more results. We find $237$ domains to be reported as Malicious and $69$ as Suspicious by, at least, one of the Security Vendors consulted by the platform. $485$ are not existing in the Virus Total database, while the remaining $2\,722$ are present but not flagged by the platform.

Using \texttt{Whois}, we check the owners of those candidates that are not flagged. We find several examples owned by companies known to be proxy domain register services like: Anonymize, Inc; Domains By Proxy, LLC; and PrivacyProtect.org. These services are a common practice for squatting domains via parking or hiding the actual identity of the domain owner. Manual checking some of those, we indeed observe a large majority of parking domains. 

The information from \texttt{Whois} let us automatically classify $37$ cases as Target-Owned domains. We observe companies such as Microsoft Corporation, Meta Platforms, Netflix, and TripAdvisor that proactively register some possibly confusing names. For instance, we found 16 domains among those we generate to be owned by Microsoft, including \texttt{mikrosoft.com} and \texttt{sharepointe.com}. Table~\ref{tab:examples_target_owned} show some examples for other brands. Appendix \ref{sec:target-owned-appendix} completes the list and provides additional details.
%while Table~\ref{tab:full_set_target_owned} in the Appendix provides the complete list. 
Notice how \TOOL is able to include extra symbols, e.g., the ``-'' in \texttt{share-point.com} that represents a silence in the compound work. We also find $16$ possible Target-Owned, but we do not have enough data to support the assumption. We list these domains in Table~\ref{tab:full_set_target_owned_doubt}.

\begin{table}[]
    \centering
    \caption{Classification of candidate Domain names that exist.}
    \label{tab:summary_second_domain_analysis}
    \begin{tabular}{c|c|c|c}

                 & \textbf{Virus Total} & \textbf{Safe Browsing} & \textbf{Whois} \\ \hline
    Not Found    & $485$                  & $-$      & $ -     $         \\ \hline
    Target-Owned & $-$                    & $-$      & $ 37    $         \\ \hline
    Malicious    & $237$                  & $4$      & $ -     $         \\ \hline
    Suspicious   & $69$                   & $-$      & $ -     $         \\ \hline \hline
    Unknown      & $2\,738$               & $3\,693$  & $ - $            \\
    \end{tabular}

\end{table}

\begin{table*}[]
\centering
\caption{Brands that proactively register confusing domains to increase the reputation of the brand and avoid squatting.}
\label{tab:examples_target_owned}
\begin{tabular}{l|l|l|c}
    \textbf{Target} & \textbf{Example}   & \textbf{Owner}        &  \textbf{\# Owned Domain} \\ \hline
        microsoft   & mikrosoft.com      & Microsoft Corporation &  7 \\ \hline
        sharepoint &  share-point.com    & Microsoft Corporation &  9 \\ \hline
        netflix     & netflics.com       & Netflix, Inc.         &  4 \\ \hline
        tripadvisor	& trypadvisor.com    & TripAdvisor LLC	     &  4 \\ \hline
        facebook    & facebuk.com        & Meta Platforms, Inc.  &  1 \\ \hline
\end{tabular}

\end{table*}

\subsection{Sound-Squatting at Python Packages}

The Python Package Index (PyPI) is a software repository for the Python programming language. PyPI allows users to search packages by name, keywords, and metadata. Several package managers use PyPI as the primary source for packages and dependencies. The ease offered by these package managers makes them susceptible to squatting, e.g., an attacker could register malicious packages with confusing names to impersonate popular packages. This attack has already happened in the past. For instance, \texttt{jeIlyfish} and \texttt{python3-dateutil} impersonated the \texttt{jellyfish} and the \texttt{dateutil} packages, respectively. They use typo- and combo-squatting, respectively. If installed, the malicious versions would steal SSH and GPG keys and send them to a remote server. Both packages have been removed from the PyPI repository.

% https://snyk.io/blog/malicious-packages-found-to-be-typo-squatting-in-PyPI/
% https://github.com/dateutil/dateutil/issues/984

An attacker could also use sound-squatting to trick users into downloading malicious packages. We investigate the possibility of the attack by collecting popular package names, generating candidates for each of them, and verifying the existence of these candidates in the official PyPI repository. The mere existence of a candidate does not prove that the package is malicious \--- whose actual identification would be cumbersome and out of the scope of this work. However, it raises some concerns about the exposure of the original package to sound-squatting. Here we focus on some interesting cases.

We consider the packages with the highest number of downloads from the official PyPI repository.\footnote{The number of downloads is not an indisputable metric. However, it is enough for this analysis.} Qualitatively, these package names present complicated patterns with dashes, version numbers, and abbreviations that challenge the generation of candidates. Yet, \TOOL can generate qualitatively good-quality candidates without the need to specialize it to the specific use case.

\begin{table}[]
    \centering
    \caption{Summary Python Package analysis.}
    \label{tab:summary_PyPI_generation}
    \begin{tabular}{c|c}
        &  \textbf{Quantity} \\ \hline
        Original Packages & $4\,299$ \\ \hline
        & $177\,998$ \\ Generated Candidates & $Mean=41.40$ \\ & $SD=19.41$ \\ \hline
        Existing Candidates & $1\,176$ \\
    \end{tabular}

\end{table}

For this analysis, we select the $5\,000$ most downloaded packages. After removing the packages with more than 20 characters,\footnote{This is the largest length our model currently accepts.} we count $4\,299$ unique packages. We run the candidate generation for each package and obtain $177\,998$ candidate names, which we verify if they exist at the PyPI repository.

As shown in Table~\ref{tab:summary_PyPI_generation}, we found $1\,176$ candidates that exist (about 0.66\% of the total of candidates). Of the $4\,299$ original packages, $719$ of them ($16.72\%$) have at least one alternative in the \TOOL candidates, which might generate confusion amount users. Table~\ref{tab:PyPI_candidates} gives some examples. By manual inspection, we find some empty projects that may become malicious in the future (as for parking domains), some joke packages, and many legitimate packages.

In a nutshell, the presence of a package does not indicate maliciousness. However, it exposes a threat that needs to be further considered. For instance, when checking some statistics that correlate to the quality of the candidate packages, we find that $522$ existing candidates have 0 stars and 0 forks.
These packages are good starting points for verification, and the maintainers of the target projects could evaluate if they are victims of sound-squatting. Similarly, tools using PyPI could warn a user when she tries to install one of the possible sound-squatted packages. \TOOL offers automatic means to generate such cases.

\begin{table}[]
    \centering
    \caption{Examples of PyPI candidates present in the repository.}
    \label{tab:PyPI_candidates}
    \begin{tabular}{c|c|l}
        \textbf{Target}                        &  \textbf{Candidate} & \textbf{Class}\\ \hline
        \href{https://PyPI.org/project/flask}{flask}   &  \href{https://PyPI.org/project/flasque}{flasque} & Empty project\\ \hline
        \href{https://PyPI.org/project/pandas}{pandas}   &  \href{https://PyPI.org/project/panndas}{panndas} & Joke\\ \hline
        \href{https://PyPI.org/project/quandl}{quandl} &  \href{https://PyPI.org/project/kwandl}{kwandl} & Legitimate \\ \hline
        \href{https://PyPI.org/project/sphinx}{sphinx} &  \href{https://PyPI.org/project/sfinx}{sfinx}   & Possible joke\\ \hline
        \href{https://PyPI.org/project/unify}{unify}   &  \href{https://PyPI.org/project/uniphi}{uniphi} & Empty project \\
    \end{tabular}
\end{table}

\section{Related Work}
\label{sec:related_work}

Nikiforakis et al. present the first work that uncovers sound-squatting as a domain name phishing technique~\cite{nikiforakis2014soundsquatting}. They propose a tool for generating sound-squatting candidates that uses a static database of homophones for English. They split the domain names into words and replace the words with homophones. After generating candidates for a list of popular domains, the authors exhaustively evaluate and categorize the candidates. The authors have also registered sound-squatting candidates, evaluating the attracted traffic to prove that sound-squatting is a possible attack vector. With \TOOL, we build on AI techniques to automatically create sound-squatting, obtaining a much more extensive list of candidates. We compare a simplified version of this work \footnote{Since no source code was provided} with our model at the Section\ref{sec:validation_results}.
%Furthermore the usage of AI paves the road to cross-language cases.

In~\cite{valentim2022ai_anon} the authors introduce an initial AI-based sound-squatting generation approach. They train a Transformers Neural Network to translate from English to IPA, introduce some noise in the process, and translate back from IPA to English to obtain sound-squatting candidates. \TOOL improves this initial idea. First, it avoids the explicit introduction of noise, which could bias the quality of the generated candidates. Second, \TOOL avoids the translation to/from IPA and builds on the actual pronunciations with the Transformers architecture. Such changes make \TOOL a general approach applicable to any language and cross-language sound-squatting scenarios.

Considering phoneme-based squatting, Sonowal and Kuppusamy focus on visually impaired people, typo-squatting, and phoneme-based phishing~\cite{sonowal2019mmsphid}. Their motivation is that visually impaired people use software to read the screen and interact with the system. They propose a system to detect phoneme-based phishing domains within the accessibility interface. In a second work, Sonowal evaluates the exposure of visually impaired people to phoneme-based phishing in email phishing campaigns~\cite{sonowal2020model}. He proposes a manual and simplistic technique to produce sound-alike keywords, using these keywords to detect phishing emails.

Considering speech, Kumar et al. uncover the \textit{skill squatting} attack, where the attacker leverages systematic errors to route users of Alexa Smart Speakers to malicious applications~\cite{kumar2018skill}. They show that around 33.3\% of the systematic errors in the speech-to-text system are due to homophones. Later, Zhang et al. formalize the skill squatting attack, calling it Voice Squatting Attack (VSA) and Voice Masquerading Attack (VMA)~\cite{zhang2019dangerous}. They also evaluate the feasibility of such attacks by deploying a malicious skill, which has been invoked by $2\,699$ users in a month.

\TOOL is orthogonal to all these efforts. It offers the automatic means to generate sound-squatting candidates. By combining the Transformers models with sound models, \TOOL works both in the same- and cross-language cases, widening the possibility of a proactive defense against sound-squatting attacks.

AI approaches, such as Transformers Neural Networks, have been used to generate artificial content such as text~\cite{Sanh2019text}, images~\cite{jiang2021transgan, pmlr-v80-parmar18a} and music~\cite{huang2018music}. These previous efforts however focus on legitimate content, where the goal is to produce realistic and correct samples. \TOOL instead aims at generating content that can confuse people. Authors of~\cite{mordvintsev2015inceptionism} \--- focusing on image generation \--- propose a way to create samples that can be understood but are perceptively confusing to people. Analogously, \TOOL generates words that are perceptively confusing regarding their pronunciation. In other words, we want our generation to induce perception errors and rely on a novel approach that introduces errors at a sub-word level.

\section{Conclusion}
\label{sec:conclusions}

In this work, we introduced \TOOL, an AI-powered sound-squatting generator. It uses Transformers Neural Networks trained to model both the phoneme representation and the pronunciation of words, producing high-quality homophones, quasi-homophones, and novel words with similar pronunciation. Because \TOOL generates sound-squatting candidates at the sub-word level, it increases the coverage while still allowing one to control the number and quality of the candidates.

We also laid the ground for cross-language sound-squatting search and mitigation. We showed that \TOOL could natively work not only with multiple languages but also cross-languages. Finally, we thoroughly analyzed \TOOL in two contexts, showing that sound-squatting is likely exploited for domain squatting already. We showed that multiple popular PyPI packages are exposed to squatting attack.

Considering in more detail the domain names case study, we found a significant number of malicious and suspicious domains already reported at Virus Total  while revealing a surprising lack of coverage for Google Safe Browsing. We also found a reasonable amount of Target-Owned domains preemptively registered by companies concerned about their reputation. In the remaining analyzed domains, we found high occurrences of domains registered in proxies, indicating some attempts to obfuscate the property. %In the package context, the finding supports the reasoning that such platforms are susceptible to sound-squatting.

For future work, we believe there is potential to increase the coverage for other applications, such as sound-squatting for smart-speakers routines, applications on app stores, and profiles on social networks. We also think the \TOOL has the potential to increase the depth of analysis and go beyond the surface of the use cases already explored.

Another possible angle of improvement is the cross-language sound-squatting, which needs a uniform approach and a thorough evaluation of its usage, including the case where smart speakers and other Speech-to-Text solutions leverage it.    

\bibliographystyle{ACM-Reference-Format}
\bibliography{bibliography}

\appendix
\section{Architectural Details}
\label{ap:details_training}
\begin{figure}[h]
    \centering
    \includegraphics[width=\linewidth]{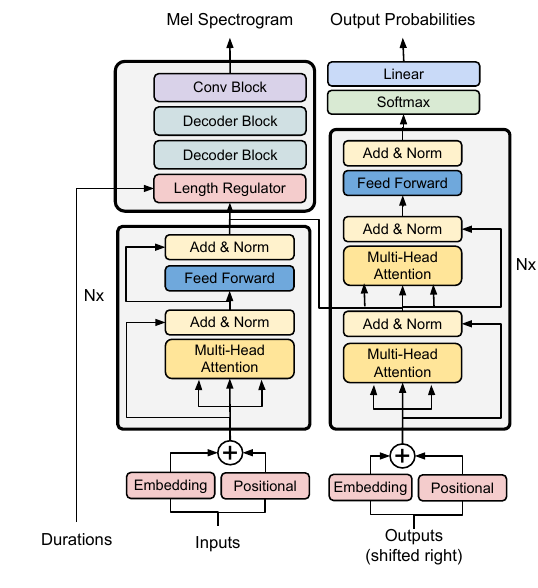}
    \caption{Full architecture of training}
    \label{fig:detailed_arch}
\end{figure}

\begin{figure}[]
%   \begin{subfigure}[b]{0.5\textwidth}
%      \includegraphics[width=\textwidth]{images/detailed_arch.pdf}
%      \caption{}
%      \label{fig:detailed_arch}
%  \end{subfigure}
%   \hspace{0.2cm}

  \begin{subfigure}[b]{0.23\textwidth}
     \centering
     \includegraphics[width=\textwidth]{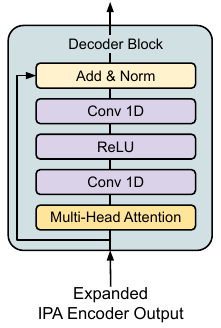}
     \caption{}
     \label{fig:decoder_block}
  \end{subfigure}
  \hspace{0.12cm}
  \begin{subfigure}[b]{0.23\textwidth}
     \centering
     \includegraphics[width=\textwidth]{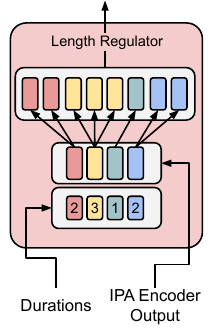}
     \caption{}
     \label{fig:length_regulator}
 \end{subfigure}
    \caption{
    (\subref{fig:decoder_block}) Inside view of the Decoder Block;
    (\subref{fig:length_regulator}) High-level representation of the Length Regulator.
    }
\end{figure}

Figure~\ref{fig:detailed_arch} details the architecture of the model used for homophone generation. We include in the diagram the \ipaencoder, \decoderTOgrapheme and \decoderTOmel. We also show in Figure~\ref{fig:decoder_block} the inside of the decoder block that we used to reduce the complexity of the visualization. Figure~\ref{fig:length_regulator} is a high-level representation of the Length Regulator. The Length Regulator expands the \ipaencoder output to the same order of magnitude as the Mel Spectrogram. This feature reduces training complexity and time. Table~\ref{tab:hyperparameters} lists the hyper-parameters.

\begin{table}[]
\centering
\caption{Compilation of hyper-parameters for \TOOL}
\label{tab:hyperparameters}
\footnotesize
\begin{tabular}{c|c}
\textbf{Hyper-parameter}                                      &   \textbf{Value}  \\ \hline
\makecell{\# Phoneme Tokens}                          & 73               \\ \hline
\makecell{\# Grapheme Tokens}                         & 33               \\ \hline
\makecell{Phoneme Embedding \\Dimension }                    & 512              \\ \hline
\makecell{\ipaencoder \\ \# of Attention Heads}               & 2                \\ \hline
\makecell{\ipaencoder \\ Hidden Dimension}                    & 512              \\ \hline
\makecell{\ipaencoder \\ Linear Hidden}                       & 2048             \\ \hline
\makecell{\ipaencoder \\ Dropout}                             & 0.1              \\ \hline
\makecell{\decoderTOgrapheme \\ \# of Attention Heads}        & 2                \\ \hline
\makecell{\decoderTOgrapheme \\ Hidden Dimension}             & 512              \\ \hline
\makecell{\decoderTOgrapheme \\ Linear Hidden}                & 2048             \\ \hline
\makecell{\decoderTOgrapheme \\ Dropout}                      & 0.1              \\ \hline
\makecell{\decoderTOmel\\Decoder Block\\Number Decoder Block} & 2                 \\ \hline
\makecell{\decoderTOmel\\Decoder Block\\Attention Hidden Dim} & 512               \\ \hline
\makecell{\decoderTOmel\\Decoder Block\\Conv1D \# Kernel}     & 9, 9              \\ \hline
\makecell{\decoderTOmel\\Decoder Block\\Conv1D \# Filters}    & 1024, 512         \\ \hline
\makecell{\decoderTOmel\\Conv Block\\Number of Conv1D}        & 6                  \\ \hline
\makecell{\decoderTOmel\\Conv Block\\Conv1D \# Kernel}        & 9                  \\ \hline
\makecell{\decoderTOmel\\Conv Block\\Conv1D \# Filters }      & 512                \\ \hline
\makecell{\decoderTOmel\\Conv Block Output\\Conv1D \# Kernel} & 9                  \\ \hline
\makecell{\decoderTOmel\\Conv Block Output\\Conv1D \# Filters}& 40                 \\ \hline
\makecell{Total Number \\of Parameters}                       & 60.2 M             \\
\end{tabular}

\end{table}

\section{Target-Owned sound-squatting candidates}
\label{sec:target-owned-appendix}

Table~\ref{tab:full_set_target_owned} lists the candidate domains registered by the same company owning the original domain. This is a clear sign that \TOOL can generate possible sound-squatting candidates that companies are already proactively considering for their brand protection. The Table~\ref{tab:full_set_target_owned_doubt} shows some examples of domains that are registered by proxy companies. Some are used for Target-Owned protection. Some are parking domains. Some are registered but not hosting any web page. 

\begin{table}[t]
    \centering
    \footnotesize
    \begin{tabular}{c|c|l}
         \textbf{Target} &        \textbf{Candidate} &                   \textbf{Owner} \\ \hline
       facebook &      facebuk.com &                               Meta Platforms, Inc. \\ \hline
         yandex &         yandx.ru &                                             YANDEX \\ \hline
      wikipedia &   wiki-pedia.org &                              Wikimedia Foundation. \\ \hline
        netflix &    netflicks.com &                                      Netflix, Inc. \\ \hline
        netflix &    netflicks.org &                                      Netflix, Inc. \\ \hline
        netflix &    netflicks.net &                                      Netflix, Inc. \\ \hline
        netflix &     netflics.com &                                      Netflix, Inc. \\ \hline
          naver &        navor.com &                                        NAVER Corp. \\ \hline
      microsoft &   microsopft.com &                              Microsoft Corporation \\ \hline
      microsoft &    mikrosoft.com &                              Microsoft Corporation \\ \hline
      microsoft &   microsofte.com &                              Microsoft Corporation \\ \hline
      microsoft &   microsofte.net &                              Microsoft Corporation \\ \hline
      microsoft &   michrosoft.com &                              Microsoft Corporation \\ \hline
      microsoft &   mikerosoft.com &                              Microsoft Corporation \\ \hline
      microsoft &   mikerosoft.net &                              Microsoft Corporation \\ \hline
          quora &        kwora.com &                                         Quora, Inc \\ \hline
          quora &       quorea.com &                                         Quora, Inc \\ \hline
     sharepoint &  share-point.com &                              Microsoft Corporation \\ \hline
     sharepoint &  share-point.org &                              Microsoft Corporation \\ \hline
     sharepoint &  share-point.net &                              Microsoft Corporation \\ \hline
     sharepoint &  sharepointe.com &                              Microsoft Corporation \\ \hline
     sharepoint &  sharepointe.org &                              Microsoft Corporation \\ \hline
     sharepoint &  sharepointe.net &                              Microsoft Corporation \\ \hline
     sharepoint & share-pointe.com &                              Microsoft Corporation \\ \hline
     sharepoint & share-pointe.org &                              Microsoft Corporation \\ \hline
     sharepoint & share-pointe.net &                              Microsoft Corporation \\ \hline
         target &      targett.org &                                Target Brands, Inc. \\ \hline
         target &      taarget.org &                                Target Brands, Inc. \\ \hline
    tripadvisor &  tripadviser.com &                                    TripAdvisor LLC \\ \hline
    tripadvisor &  tripadviser.org &                                    TripAdvisor LLC \\ \hline
    tripadvisor &  tripadviser.net &                                    TripAdvisor LLC \\ \hline
    tripadvisor &  trypadvisor.com &                                    TripAdvisor LLC \\ \hline
           hulu &        hoolu.com &                                          Hulu, LLC \\ \hline
           hulu &        hoolu.org &                                          Hulu, LLC \\ \hline
           hulu &        hoolu.net &                                          Hulu, LLC \\ \hline
           hulu &        huloo.org &                                          Hulu, LLC \\ \hline
           hulu &        huloo.net &                                          Hulu, LLC \\  
    \end{tabular}
    \caption{Full compilation of found Target-Owned domains.}
    \label{tab:full_set_target_owned}
\end{table}

\begin{table}[t]
    \centering
    \footnotesize
    \begin{tabular}{c|c|l}
         \textbf{Target} &        \textbf{Candidate} &                   \textbf{Owner} \\ \hline
         
       spankbang &    spankbank.org &                               Redacted for Privacy \\ \hline
       stripchat &   stripchatt.com &                               Redacted for Privacy \\ \hline           
       pornhub   &     pornhubb.com &                                      Whois Privacy \\ \hline
         pornhub &     porn-hub.org &                                      Whois Privacy \\ \hline
         pornhub &      pornhab.com &                                      Whois Privacy \\ \hline
     wildberries &    wildberris.ru &                                     Private Person \\ \hline
     wildberries &    wildberies.ru &                                     Private Person \\ \hline
           zhihu &       zhihoo.com &                               Redacted for Privacy \\ \hline
           deepl &      deepaul.com &                               Redacted for Privacy \\ \hline
       onlyfans &    onlyphans.net &                              Domains By Proxy, LLC \\ \hline
       telegram &    telegramm.org &                              Domains By Proxy, LLC \\ \hline
       telegram &   telegramme.org &                              Domains By Proxy, LLC \\ \hline
       telegram &    tellegram.com &                              Domains By Proxy, LLC \\ \hline
       telegram &    telligram.com &                              Domains By Proxy, LLC \\ \hline
       flipkart &     flipcart.net &                              Domains By Proxy, LLC \\ \hline
       flipkart &     flypkart.com &                              Domains By Proxy, LLC \\ 
    \end{tabular}
    \caption{Other Target-Owned compilation, however owned by proxy and private individuals.}
    \label{tab:full_set_target_owned_doubt}
\end{table}

\end{document}